\documentclass[aps,prl,reprint,groupedaddress]{revtex4-2}
\usepackage{graphicx}
\usepackage{color}
\usepackage{ulem}
\usepackage{lineno}
\usepackage[colorlinks=true,
linkcolor=blue,
citecolor=blue,
urlcolor=blue]{hyperref}
\newcommand{\degree}{$^{\circ}$}
\begin{document}


\title{Light-induced spin-polarized desorption of Rb atoms from Co surfaces}


\author{Kanta Asakawa$^{1,2}$}
\email[]{asakawa-kanta-hn@ynu.ac.jp}
\author{Naoki Tanabe$^2$}
\author{Oki Watanabe$^3$}
\author{Shuji Kamada$^3$}
\author{Keisuke Hara$^3$}
\author{Kaori Niki$^3$}
\author{Atsushi Hatakeyama$^2$}
\affiliation{$^1$Department of Physics, Graduate School of Engineering Science, Yokohama National University, Yokohama 240-8501, Japan}
\affiliation{$^2$Department of Applied Physics and Chemical Engineering, Tokyo University of Agriculture and Technology, Koganei, Tokyo 184-8588, Japan}
\affiliation{$^3$Graduate School of Science, Chiba University, Chiba 263-8522, Japan}

\date{\today}

\begin{abstract}
 \textcolor{black}{We investigated the spin polarization of Rb atoms undergoing light-induced desorption from a spin-polarized Co (110) surface. Desorption of Rb atoms was induced via pulsed UV irradiation of the Rb-adsorbed Co surface. The number of desorbed atoms, their velocity distribution, and their spin polarization were measured via the absorption of circularly polarized light resonant with the desorbed Rb atoms. The results were analyzed in conjunction with DFT calculations, indicating that the desorbed Rb atoms were spin-polarized. This was explained by a non-thermal desorption mechanism accompanying spin-polarized electron transfer between the Co surface and adsorbed Rb.}  
\end{abstract}


\maketitle


Interactions between gaseous atoms and magnetic surfaces often change the spin states of both phases. 
For example, charge transfer between a magnetic surface and adsorbates often modifies the magnetic properties of the surface. Examples include demagnetization\cite{PhysRevLett.39.568}, changes in the magnetic moments of surface atoms\cite{cao2022spin,cao2023spin,asakawa2019electronic}, and the enhanced spin polarization of conduction electrons\cite{kurahashi,parkinson}.  Charge transfer may also occur between the surface and adsorbates during desorption. When the surface electrons are spin-polarized, the transferred electrons are also expected to be spin-polarized, leading to spin polarization of the desorbed atoms. This may be  important when ferromagnetic catalysts promote oxygen evolution reactions\cite{ren2021spin,fang2023spin}. However, spin transfer during desorption has not been investigated sufficiently due to the difficulty of detecting spin transfer  between the surface and adsorbates experimentally. In the case of desorption from spin-polarized surfaces, such as ferromagnets, detection of spin polarization of the desorbed atoms provides direct evidence for spin transfer between the surface and  adsorbates during desorption. In non-thermal fast desorption processes, such as light-induced desorption\cite{ikeda}, desorption should occur before spin relaxation occurs; therefore, spin polarization of the desorbed atoms should be readily observable.
   
In this study, we observed the spin polarization of atoms subjected to light-induced desorption from a ferromagnetic surface using a method that exploits the spin-selective absorption of circularly polarized light\cite{asakawa2023optical}. Co  and Rb served as the ferromagnet and 
  the adsorbate, respectively. Adsorption of alkali metal atoms onto Co surfaces is accompanied by both charge transfer and changes in the Co atom spins\cite{cao2022spin,cao2023spin,zhang2024spin}. Therefore, adsorption of Rb onto Co is likely associated with spin-polarized electron transfer. We found that on irradiation with pulsed UV light, Rb atoms desorbed from  a Co surface via a non-thermal mechanism, and the desorbed atoms showed spin polarization, indicating that the desorption process involves spin-polarized charge transfer.  

\begin{figure}
	\centering
	\includegraphics[width=0.7\linewidth]{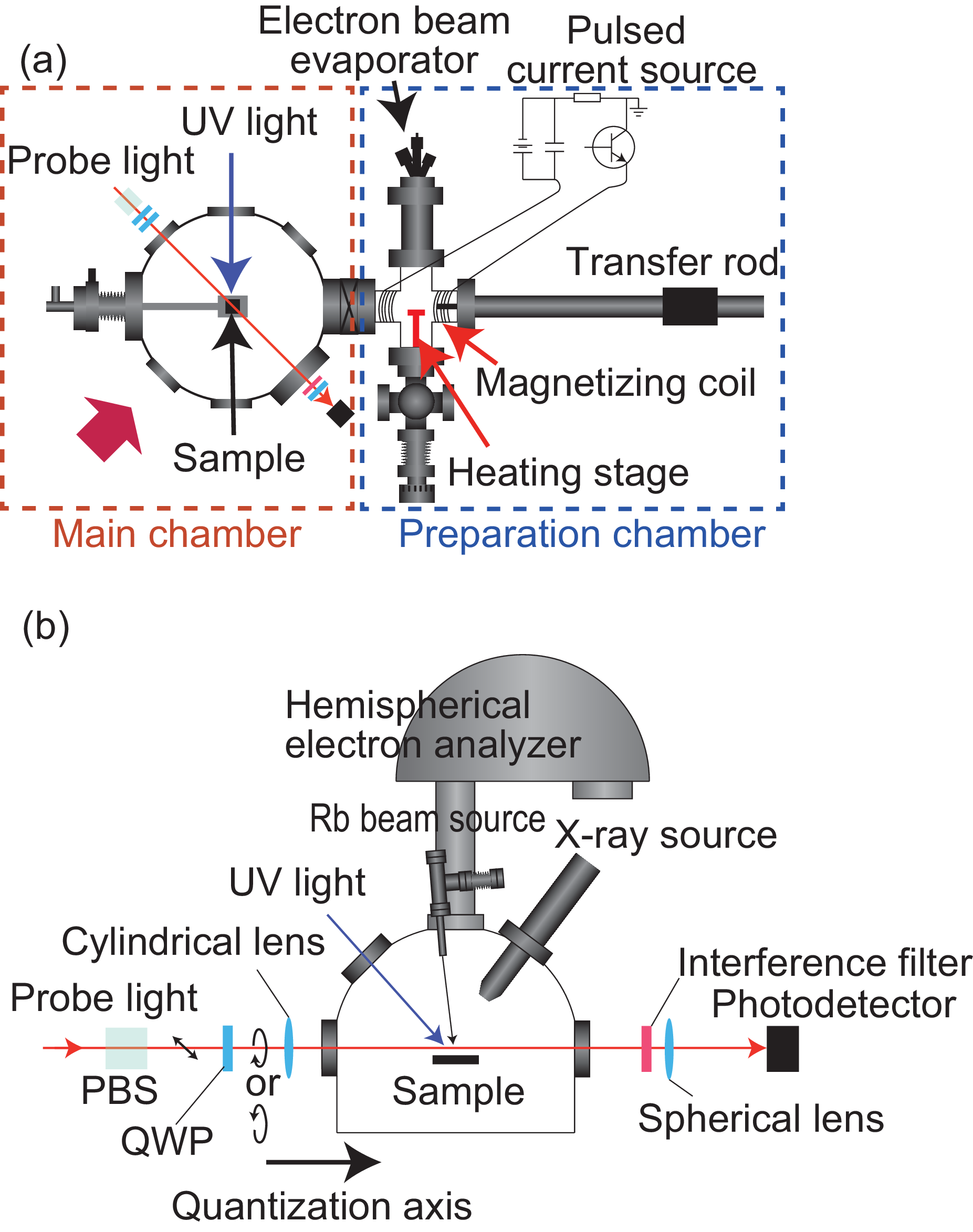}
	\caption{(a) Top view of the apparatus and (b) side view of the main chamber viewed from the direction shown by the red arrow in (a). }
	\label{fig:liadapparatus}
\end{figure}

The experimental apparatus is shown in Fig. \ref{fig:liadapparatus}(a). Two ultra-high-vacuum (UHV) chambers—preparation and measurement chambers—were separated by a gate valve. The base pressures were approximately $3\times 10^{-6}$ Pa and $2\times 10^{-7}$ Pa, respectively. The preparation chamber featured a transfer rod, a sample heating stage, and an electron-beam evaporator. The measurement chamber was equipped with a sample stage, viewports for the probe  and  UV lights, a multi-channel effusive atomic Rb beam source, an X-ray source and a hemispherical electron analyzer required for X-ray photoelectron spectroscopy (XPS). \textcolor{black}{To shield external magnetic fields such as the Earth's magnetic field, the measurement chamber was made of permalloy.}
Details of the Rb beam source were described in Ref. \cite{asakawaPRA}. The flux intensity was estimated to be $\sim10^{13}$ atoms per second based on the designed value.
The probe light and pulsed UV light were introduced through the viewports. The UV light was generated by the third-harmonic generation of a Nd:YAG laser. The wavelength, pulse width, repetition rate, 1/e$^2$ half width, and fluence were 355 nm, 5–10 ns, 10 Hz, 7.0 mm, and 65.0 mJ/cm$^2$, respectively. The angle of incidence of the UV light on the sample was approximately 45$^{\circ}$, and the light was s-polarized.

The probe light was generated by an external-cavity diode laser.  Its frequency was tuned to the $F=3$ $\rightarrow$ $F'=4$ transition frequency of the \textcolor{black}{$^{85}$Rb} $D_2$ line using polarization spectroscopy, where $F$ and $F'$ are the total angular momenta of Rb atoms in the $5^2{\rm S}_{1/2}$ ground state and the $5^2{\rm P}_{3/2}$ excited state, respectively.  Figure \ref{fig:liadapparatus}(b) shows a side view of the main chamber and the optical system for the probe light. The horizontally polarized component of the probe light was extracted by a polarizing beam splitter (PBS) and then converted into circularly polarized light by a quarter-wave plate (QWP) mounted on a piezo-driven rotation mount. The helicity of  circular polarization was controlled by rotating the QWP. A cylindrical lens was then used to focus the circularly polarized probe light   vertically to a point 1.0 mm above the sample. \textcolor{black}{To ensure that the effect of the chamber windows on the purity of the circular polarization of the probe light was small, we placed a mirror on the far side of the chamber to reflect the probe light back through the two vacuum chamber windows, the cylindrical lens, and the quarter-wave plate toward the PBS. We then measured the intensity of the light transmitted through the PBS. Ideally, after passing through the quarter-wave plate twice, the probe light should be converted from horizontal to vertical polarization and should not be transmitted through the PBS. The fraction of light transmitted through the PBS was measured to be 3 \% of the total intensity, confirming that the chamber windows had a minimal influence on the circular polarization state of the probe light.} The helicity of  circular polarization was switched every 20 UV pulses. \textcolor{black}{The probe light intensity was maintained below 8 \textmu W to avoid influencing the spins of the desorbed atoms. Before measurements, we confirmed that changing the helicity altered the probe light intensity by less than 1 \%, ensuring that variations in probe intensity would not introduce systematic errors in the measured spin polarization of the desorbed atoms.}

 After passing through the chamber, the probe light was focused by a convex lens and then entered a photodetector. An interference filter   between the chamber and the lens ensured that scattered UV light did not reach the photodetector. The spin polarization of the desorbed Rb atoms was estimated from the dependence of probe light absorbance on the helicity of the circular polarization of the probe light. Note that the ground states of Rb include two hyperfine states, $F=2$ and $F=3$, with the former being 3,036 MHz (12.56 \textmu eV) lower than the latter. The probe light detected only $^{85}$Rb atoms in the $F=3$ state. 
\textcolor{black}{Before the experiment, we validated the effectiveness of our method by measuring the spin polarization of Rb atoms desorbed from a platinum surface and subsequently spin-polarized by circularly polarized pump light at the same frequency as the probe light (data not shown).}

An epitaxial  face-centered cubic (fcc)  Co thin film was grown on an MgO (110) substrate\cite{nukaga2009structure} that had been cleaned in the preparation chamber by annealing at  600 \degree C  for approximately two hours in UHV. The cleanliness of the substrate was confirmed by XPS. A 10-nm Co thin film  was then deposited via using electron beam evaporation at a deposition rate of 0.041--0.067 \AA/s. During deposition, the MgO (110) substrate was held at 500 \degree C. The Co film thus  prepared exhibited an fcc (110) structure, as confirmed by reflection high-energy electron diffraction (RHEED) using another apparatus. The sample was then magnetized along the [001] easy axis by a coil  wound 200 times around the preparation chamber. A pulsed current source delivered 360 A, which generated a magnetic field of 1.7 kOe, as measured by a pick-up coil. This adequately magnetized the sample, as confirmed by the magnetization curve obtained using a superconducting quantum interference device (SQUID).
The Co (10 nm)/MgO sample was then transferred to the main chamber and mounted on the sample stage, with the directions of magnetization and the probe light aligned. XPS was used to confirm that the Co film was clean. Measurements proceeded at room temperature.

To gain deeper insights into the electronic structure and desorption mechanism, we performed density functional theory (DFT) calculations. DFT calculations were performed using the Vienna ab initio simulation package (VASP), ver. 6.4.1 \cite{43,44}. Exchange correlation effects were described using the spin-polarized generalized gradient approximation (GGA) within the framework of the revised Perdew-Burke-Ernzerhof (RPBE) formalism \cite{22}. A plane-wave basis set was used in conjunction with the projector augmented wave (PAW) \cite{45} method, allowing a relatively low kinetic energy cut-off of about 400 eV. Brillouin-zone integration was conducted performed using a Monkhorst-Pack $3 \times 3 \times 1$ grid of k-points grid \cite{47}. The energy convergence criterion for self-consistency cycles was established at $1\times 10^{-8}$ eV. Optimization of the ionic positions continued until the residual forces on each ion had been reduced to below $1\times 10^{-4}$ eV\AA$^{-1}$. We incorporated such treatments into another set of calculations with the GGA functional using the zero-damping DFT-D3 method of Grimme et al. \cite{53}.
We modeled the Rb/Co(110) surface using a Co(110) slab composed of five atomic layers of Co in a $(6\times6)$ surface unit cell with and a vacuum region of 25 \AA$ $ to avoid spurious interactions between periodic images. For bulk fcc Co, a lattice parameter of 3.59 \AA$ $ was obtained, which was in accordance with the experimental value of 3.55 \AA \cite{42}. This structure has a magnetic moment of 1.6 \textmu$_\mathrm{B}$ per Co atom, where \textmu$_\mathrm{B}$ is the Bohr magneton. For Rb adsorption, we considered the short-bridge site and fourfold hollow site on the Co(110) surface. The adsorption energy ($E_{\rm{ads}}$) was calculated using the following equation; 
\begin{equation}
	E_{\rm{ads}} = E_{\rm{mol/slab}}- E_{\rm{mol}} - E_{\rm{slab}}
\end{equation}
where $E_{\rm{mol/slab}}$, $E_{\rm{mol}} $, and $E_{\rm{slab}}$ are the total energy of the slab with the adsorbate, the total energy of an isolated molecule in the gas phase, and the total energy of the slab, respectively.


\begin{figure}
	\centering
	\includegraphics[width=0.7
	\linewidth]{"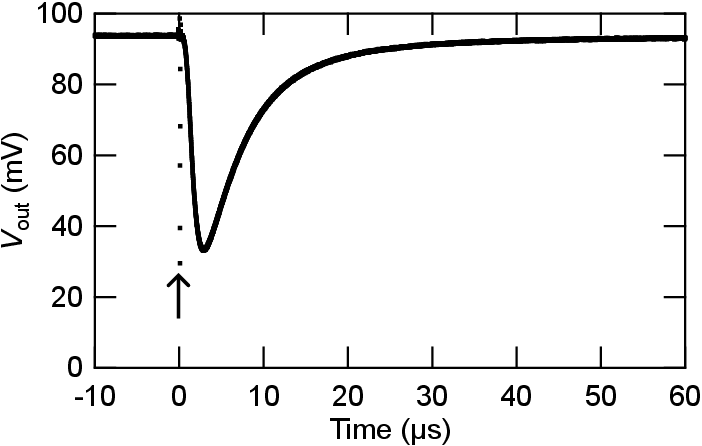"}
	\caption{Time dependence of photodetector output voltage showing the absorption of probe light by desorbed atoms after UV irradiation.}
	\label{fig:tof-spectrum}
\end{figure}

Figure \ref{fig:tof-spectrum} shows a representative time spectrum of the photodetector output voltage. The noise  at the time origin (indicated by the arrow) originates from the UV light. The dip following UV light irradiation is attributed to the absorption of the probe light by desorbed Rb atoms. We define the absorbance $A(t)$ as
\begin{equation}
	A(t)=\ln\left(\frac{V_0}{V_{\rm out}(t)}\right),
\end{equation} 
where $V_{\rm out}(t)$ is the output voltage of the photodetector and $V_0$ is the baseline of the output voltage obtained by averaging $V_{\rm out}(t)$ over the region $-60\rm{ \textmu s} <t<-0.15\textrm{ \textmu s} $. As the absorbance is proportional to the density of desorbed atoms that are resonant with the probe light, the number of desorbed atoms is proportional to the product of $A$ and the velocity of the atoms. Therefore, we introduce  a ``quantity'' $I_{\rm des}$, defined below, which is proportional to the number of desorbed atoms:

	\begin{eqnarray}
	I_\mathrm{des}&=&\int_{t_1}^{t_2}A(t) v_{z} (t)dt\nonumber\\ 
	&=&\int_{t_1}^{t_2}A(t)\cdot \frac{ l}{t} \mathrm{d}t,
\end{eqnarray}
where $v_{z}$ is the velocity component along the surface-normal direction and $l$ is the distance between the sample and the probe light, which was $1.0\times 10^{-3}$ m. $t_1=0.47$ \textmu s and $t_2=40.47$ \textmu s were adopted, which covers the entire peak and excludes the UV-laser-derived noise. Note that $A(t)$ measures the density of resonant atoms, which in turn means that the flux of desorbed atoms is proportional to $A(t)v_z(t)$.

\begin{figure*}
	\centering
	\includegraphics[width=0.7\linewidth]{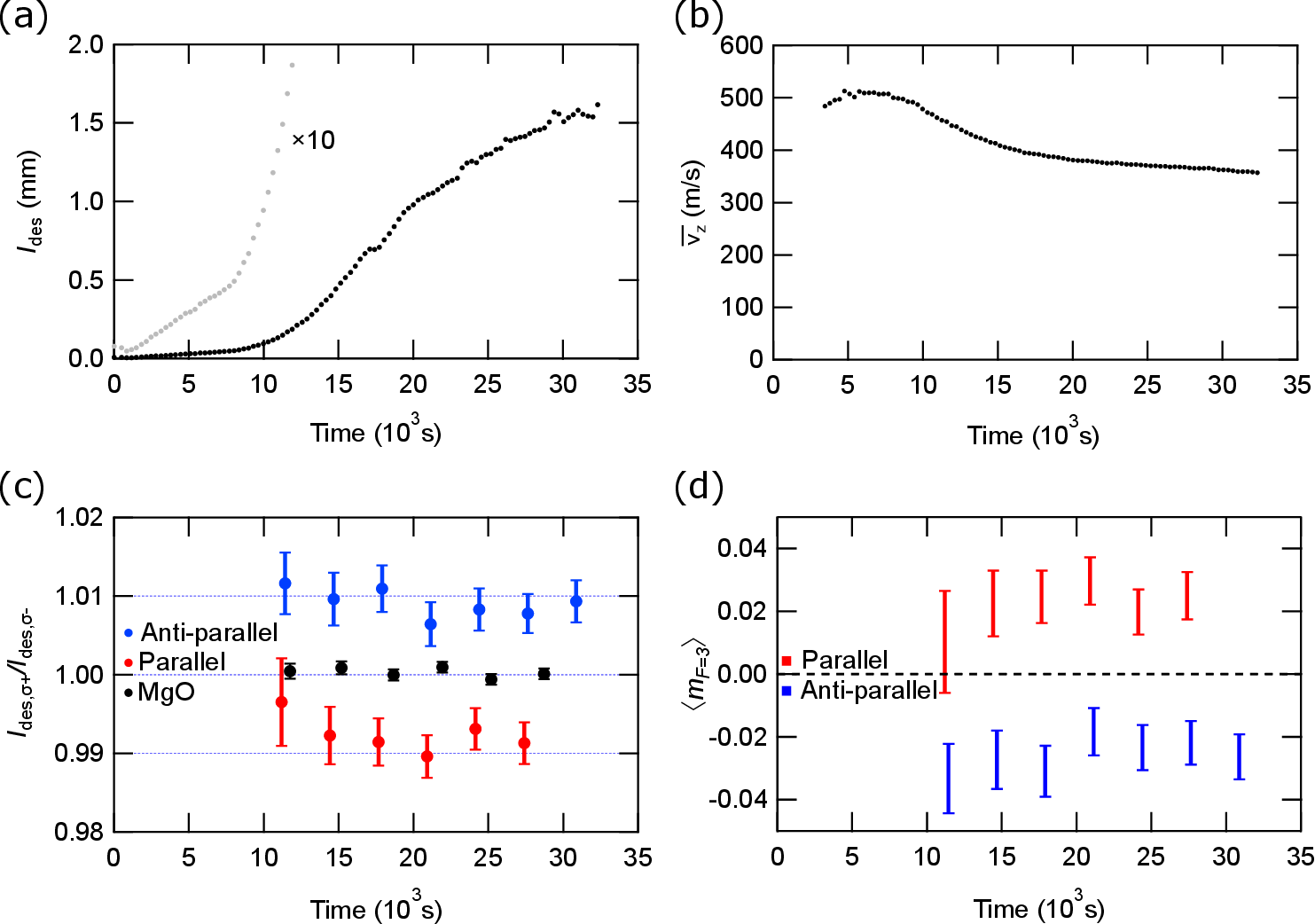}
	\caption{The time evolution of (a) the desorption intensity $I_{\rm des}$, (b) the  mean velocity component along the surface-normal direction, (c)  $I_{\rm des,\sigma +}/I_{\rm des,\sigma -}$, and (d) the averaged magnetic quantum number. The time origin corresponds to the start of UV and Rb beam irradiation.}
	\label{fig:polarization}
\end{figure*}

 Figure \ref{fig:polarization}(a) shows the time evolution of $I_{\rm des}$. For each data point, $I_\mathrm{des}$ was calculated from the spectra averaged over 2,000 UV laser pulses. The right axis indicates the time from the start of Rb beam and UV irradiation.  Contrary to the case of Rb desorption from Fe$_3$O$_4$, which exhibited threshold coverage below which desorption was not detected\cite{asakawa2023optical}, desorption commenced immediately after  Rb irradiation. $I_{\rm des}$ increased linearly with time  from 0 s to $\sim 8 \times 10^3$ s. From  that time onward, the rise was steeper, implying  the presence of at least two types of desorption component with different desorption cross sections; the component of the lower desorption cross section are saturated first, after which the  higher cross-section component became dominant, explaining the steeper rise. The increase in $I_{\rm des}$ slowed gradually after $\sim 2 \times 10^4$ seconds, implying that the number of desorbed atoms per UV pulse was approaching equilibrium with the number of atoms adsorbed between UV pulses and that Rb coverage was nearing saturation, or that a sufficiently thick Rb multilayer had formed, such that the desorption rate no longer depended on the film thickness. The mean velocity component along the surface-normal direction, $\overline{v_{z}}$, was estimated using the following equation:
\begin{eqnarray}
	\overline{v_{z}}&=&\frac{\int_{t_1}^{t_2}A(t) v_{z}^2(t)dt}{\int_{t_1}^{t_2}A(t)v_{z}(t)dt}\nonumber \\
	&=&\frac{\int_{t_1}^{t_2}A(t)\left(\frac{l}{t}\right)^2dt}{I_\mathrm{des}}.
\end{eqnarray}
As shown in Fig. \ref{fig:polarization}(b),  $\overline{v_{z}}$ decreased from  500 m/s at $8\times 10^3$ s to 357 m/s at $3.2\times 10^4$ s, corresponding to the temperatures of 1,200 K and 615 K obtained by earlier Monte-Carlo simulations \cite{asakawa2023optical} that assumed a Maxwell-Boltzmann (M-B) distribution and application of the  Knudsen's cosine law. These values were significantly larger than those obtained under similar conditions using Fe$_3$O$_4$. The decrease in $\overline{v_{z}}$ also contrasted with what was observed for Fe$_3$O$_4$\cite{asakawa2023optical}, in which $\overline{v_{z}}$ increased monotonically with time; this was explained by assuming a thermal desorption mechanism. In the present case, the decrease in $\overline{v_{z}}$ with time implies that the desorption was non-thermal. A similar tendency has been reported for the desorption of K atoms from Cr$_2$O$_3$\cite{wilde1999}, on which K adsorbed as K$^+$ and the desorption was driven by light-induced excitation, which resulted in the substrate electrons being transferred to K orbitals, neutralizing the K$^+$ and causing it to desorb. The decrease in $\overline{v_{z}}$ was attributed to a change in the adsorption mechanism — from strongly ionic bonding in the low-coverage regime to weakly ionic bonding at higher coverage — with the former atoms exhibiting a higher desorption velocity than the latter. These similarities strongly imply that the desorption of Rb from the fcc-Co(110) surface observed here was also driven by electronic transition involving electron transfer between the surface and the adsorbate.

The spin polarizations of  desorbed atoms  were estimated by comparison of  $I_{\rm des,\sigma +}$ and $I_{\rm des,\sigma -}$ data obtained using  left ($\sigma^+$) and right($\sigma^-$) circularly polarized probe light, respectively. Figure \ref{fig:polarization}(c) shows the time evolution of $I_{\rm des,\sigma +}/I_{\rm des,\sigma -}$ after $1.0\times 10^4$ s, at which point the desorption intensity became sufficiently large to yield reliable values. For each datum,  $I_{\rm des,\sigma +}/I_{\rm des,\sigma -}$ was calculated using information from  $2\times 10^4$  UV laser shots.  The obtained  $I_{\rm des,\sigma +}/I_{\rm des,\sigma -}$ for the Co/MgO samples deviated from unity, and the direction of deviation varied with changing the magnetization direction of the sample. \textcolor{black}{This trend was consistently reproduced in a repeat experiment using another sample.} The black points are those for a  (control) MgO(110) substrate (without the Co thin film) obtained under the same conditions except that the UV laser fluence  was 97.5 mJ/cm$^2$. For the pure MgO substrate, the deviation in  $I_{\rm des,\sigma +}/I_{\rm des,\sigma -}$ from unity was within the range of experimental uncertainty. \textcolor{black}{This result, combined with the observation that the sign of the spin polarization of Rb atoms desorbed from a Co surface reverses upon changing the direction of magnetization, confirm that the observed deviation in  $I_{\rm des,\sigma +}/I_{\rm des,\sigma -}$ from unity originated from magnetization of the sample}. 

In order to demonstrate that the deviation of $I_{\rm des,\sigma +}/I_{\rm des,\sigma -}$ from unity represents the spin polarization of the desorbed atoms, it is also necessary to rule out the influence of the stray magnetic field from the sample. The Zeeman splitting induced by the stray magnetic field may lead to selective excitation of a particular magnetic sublevel by the probe light. This may also result in a deviation in  $I_{\rm des,\sigma +}/I_{\rm des,\sigma -}$ from unity\cite{budker2002}. If the edge of the sample is regarded as a line-shaped magnetic pole, the magnitude of the stray magnetic field at the position of the probe light is approximated by the following equation.
\begin{equation}
B\sim\frac{\mu_0 M_{\rm{s}}t}{2\pi l},
\end{equation}
where $B$ is the the magnetic flux density at the probe light position, $\mu_0$ is the permeability in vacuum, $M_{\rm{s}}$ is the remanent magnetization of the Co thin film, $t$ is the  thickness of the Co thin film. Here,  $M_{\rm{s}}$ was $9.63\times10^5$ A/m as measured using SQUID, and $t$ was 10 nm. Based on this equation, $B$ was estimated to be $\sim 1.9$ \textmu T. This field results in a Zeeman splitting  of $\sim2.2\times 10^{-10}$ eV \textcolor{black}{between the $m_{F=3}=-3$ and $m_{F=3}=3$ hyperfine levels}, which corresponds to a frequency of $\sim 55$ kHz.  However, this shift is  negligibly small compared  the Doppler broadening at room temperature ($\sim 100$ MHz). Therefore, under the present experimental conditions, all magnetic sublevels are equally excited by the probe light and the effect of the stray magnetic field on $I_{\rm des,\sigma +}/I_{\rm des,\sigma -}$ can be neglected.  This means that the observed deviation in $I_{\rm des,\sigma +}/I_{\rm des,\sigma -}$  from unity was attributable to spin polarization of desorbed atoms, and that such polarization depended on the direction of sample magnetization. 

The averaged magnetic quantum number $\left\langle m_{F=3}\right\rangle$ of atoms in the $F=3$ ground state was calculated using a method described in Ref. \cite{asakawa2023optical}. \textcolor{black}{Because the optical excitation cross-section of a Rb atom in the $F=3$ state under circularly polarized probe light is described by a quadratic function of $m_{F=3}$, the ratio $I_{\rm des,\sigma +}/I_{\rm des,\sigma -}$  can be expressed as a function of $\left\langle m_{F=3}\right\rangle$, $\left\langle m_{F=3}^2\right\rangle$, and the velocity distribution of the desorbed atoms. The velocity distribution of the desorbed atoms is necessary to account for atoms that are excited to states other than the $F' = 4$ state ($F'=2$ and $F'=3$ states) due to Doppler shifts. By approximating the velocity distribution of desorbed atoms by a M-B distribution with Knudsen's cosine law, we mapped the calculated  $I_{\rm des,\sigma +}/I_{\rm des,\sigma -}$ as a function of $\langle m_{F=3} \rangle$ and $\langle m_{F=3}^2 \rangle$, and compared it with the experimentally obtained values to determine the range of $\langle m_{F=3} \rangle$ consistent with the measured $I_{\rm des,\sigma +}/I_{\rm des,\sigma -}$ within its standard uncertainty.  }  The temperature of the M-B distribution was chosen to match the mean velocity along the surface-normal direction with the experimentally obtained $\overline{v_{z}}$ data. Even when desorption is non-thermal, the desorption velocity distribution can often be approximated by an M-B distribution\cite{zimmermann1994velocity}.
\textcolor{black}{In some cases, however, a single M-B distribution cannot describe the velocity distribution, due to the presence of multiple contributing components such as diffusive processes at high coverages. Nonetheless, we consider this approximation to  be sufficient for the present purpose, because varying the assumed temperature from 500 to 10,000 K changed the estimated $\langle m_{F=3} \rangle$ value by less than 0.003. This small change indicates that estimation of $\langle m_{F=3} \rangle$ is only weakly dependent on the assumed velocity distribution.}

 The results are shown in Fig. \ref{fig:polarization}(d). The error bars are the ranges of values that $\left\langle m_{F=3}\right\rangle$ may take. When the sample was magnetized along the direction parallel to that of the probe light, the desorbed atoms were spin-polarized along the same direction, indicating that the desorbed atoms were spin-polarized along the direction parallel to the direction of magnetization. As the major electron spin direction was anti-parallel to that of magnetization, the spin-polarization direction of the desorbed atoms was parallel to the minority-spin direction of the sample. The $\left\langle m_{F=3}\right\rangle$ values averaged over atoms that desorbed after $1.0\times 10^4$ s were $-0.024\pm0.003$ and $0.024\pm0.004$ for the ``anti-parallel" and ``parallel" configurations, respectively. 
  
  \begin{figure}
  	\centering
  	\includegraphics[width=0.7\linewidth]{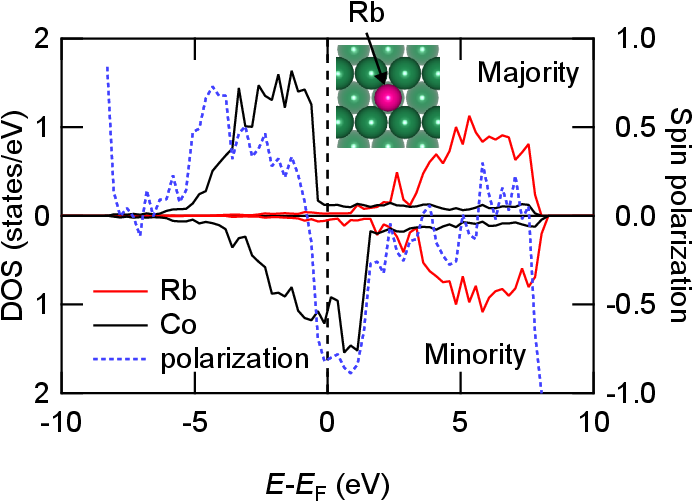}
  	\caption{The calculated DOS of the surface Co atoms and the adsorbed Rb atom.  The inset shows the adsorption site of Rb used for the calculation visualized by VESTA software\cite{Momma:db5098}.}
  	\label{fig:dos}
  \end{figure}
 
Figure \ref{fig:dos} shows the calculated density-of-states (DOS) of an adsorbed Rb atom and one of the four Co atoms closest to the Rb atom. The calculated adsorption energy $E_{\rm{ads}}$ was  $-2.02$ eV. The inset shows the adsorption site of the Rb used in the calculation. The dotted line indicates the spin polarization of the Co  DOS, defined by;
\begin{equation}
P=\frac{D_{\uparrow}-D_{\downarrow}}{D_{\uparrow}+D_{\downarrow}},
\end{equation}
where $P$ is the spin polarization, and $D_\uparrow$ and $D_\downarrow$ are the majority and minority-spin DOSs of the Co atom, respectively. In the region $-0.6$ eV$< E-E_{\rm F}\leq$ $0$ eV, the minority-spin Co $3d$ states exhibited larger DOSs than did the majority-spin state. Therefore, the DOS near the Fermi level was polarized along the minority-spin direction. However, in the region $-6.3$ eV$\leq E-E_{\rm F}\leq$ $-0.6$ eV, the DOS of the majority-spin state was larger than those of the minority-spin states. The Co 3$d$ states extended above the Fermi level but the DOSs were low. These features are in good agreement with those of Gunn \textit{et al.} \cite{GunnPRB}. The Rb $5s$ state,  which is the  outermost orbital of an isolated Rb atom, lay above the Fermi level, indicating that Rb donated its 5$s$ electrons to the substrate upon adsorption. 

As the desorbed atom spins were polarized along the minority-spin direction, it is likely that the Rb gained electrons from the Co $3d$ states near the Fermi level. One possible mechanism of desorption is that irradiation with UV light of photon energy 3.49  eV  excited Co $3d$ electrons near the Fermi level to states above the Fermi level, after which these electrons were transferred to adsorbed Rb atoms, neutralized  the Rb atoms, and the atoms then desorbed. 
If the UV light excites the electrons at the Fermi level, where the spin polarization is $-0.80$, and the excited electrons are transferred to the desorbed atoms, the observed $\left\langle m_{F=3}\right\rangle$ is expected to be 0.40 in the case of ``parallel" configuration, since the electron spin quantum number is 1/2 and the spin and magnetic moment are in opposite directions. \textcolor{black}{ This is much larger than the experimentally-obtained values.} In practice, however, excitation may occur from deeper electronic states with lower spin polarization than that at the Fermi level, and spin relaxation of desorbing atoms may occur during desorption. 
For simplicity, the calculation assumed monolayer adsorption; in practice, however, desorption of atoms from the Rb multilayer may also occur. In such cases, the desorbing Rb atoms are less likely to be spin-polarized because they do not receive spin-polarized electrons directly from the Co. Although we did not measure XPS while measuring light-induced desorption, due to instrument limitations, the Rb irradiation dose was sufficient to cause multilayer adsorption. These effects may have resulted in small $\left\langle m_{F=3}\right\rangle$ values.

Note that our method detects only $^{85}$Rb atoms in the $F=3$ state, and that the spin polarization of desorbed atoms in the $F=2$ state remains unknown. However, the amount of spin transferred to atoms that are desorbing is presumably independent of hyperfine status. Therefore, our results may well-represent the spin polarization of all desorbed atoms.
\textcolor{black}{It is important to note that DFT results do not exclude the possibility of mechanisms other than those described above. In particular, the electron and spin dynamics during the desorption process remain a matter of speculation. To clarify these in more detail, systematic studies are required, such as comparing the desorption cross sections, velocity distributions, and spin polarizations of desorbed atoms under pulsed light irradiation at various wavelengths. This will be the focus of the next stage of our research. }
 
In conclusion, the spin polarization of Rb atoms that desorbed from a spin-polarized fcc-Co (110) surface upon UV-light irradiation was investigated using an experimental method that used spin-selective optical detection of desorbed Rb atoms, and DFT calculations. \textcolor{black}{We found that Rb atoms desorbed from the Co surface is spin-polarized, which is in contrast to the case of thermal desorption  of Rb from Fe$_3$O$_4$\cite{asakawa2023optical}. This result, together with the decrease in  the desorption velocity with time, suggests that the desorption of Rb from the Co surface is driven by a non-thermal mechanism involving spin-polarized charge transfer between the surface and the adsorbates, triggering spin polarization of the desorbed atoms.} These results demonstrate that this method for measuring the spin polarization of desorbed atoms provides insights into the mechanisms of adsorption and light-induced desorption from spin-polarized surfaces. In research on surface desorption processes, the spin state of desorbed atoms has long remained unclear due to the lack of a suitable detection method\cite{HIROSE20072740}. This has posed a major obstacle to elucidating surface chemical reactions involving spins. The results presented here demonstrate that our method can detect spin transfer during desorption. Although this approach is applicable only when desorbing atoms can be optically detected, it holds significant potential to address fundamental questions on spin behavior during charge transfer in adsorption and desorption processes. Furthermore, it may elucidate spin-related surface reactions, such as the promotion of ferromagnetic catalysts by alkali metal adsorption\cite{cao2022spin,cao2023spin,zhang2024spin}. In addition, the phenomenon of light-induced spin-polarized desorption can potentially be applied to polarized atomic sources.
  
\begin{acknowledgments}
This work was supported by Grant-in-Aid for Scientific Research from the Japan Society for the Promotion of Sciences (JSPS KAKENHI) Grants No.  JP21K13864 and JP24K01343 and “Advanced Research Infrastructure for Materials and Nanotechnology in Japan (ARIM)” of the Ministry of Education, Culture, Sports, Science and Technology (MEXT) Proposal Number JPMXP1224UE5367.
\end{acknowledgments}
The data that support the findings of this article are openly available at \cite{dataset}.

\end{document}